\documentclass[aps, prl, showpacs, twocolumn, amsfonts, amsmath, amssymb, superscriptaddress, floatfix]{revtex4}

\usepackage{graphicx}
\usepackage{dcolumn}
\usepackage{bm}

\begin{document}

\title{Observation of a Cooperative Radiation Force in the Presence of Disorder}

\author{T. Bienaim\'e}
\affiliation{Institut Non Lin\'eaire de Nice, CNRS, Universit\'e de Nice Sophia-Antipolis, 06560 Valbonne, France}

\author{S. Bux}
\affiliation{Institut Non Lin\'eaire de Nice, CNRS, Universit\'e de Nice Sophia-Antipolis, 06560 Valbonne, France}
\affiliation{Physikalisches Institut, Eberhardt-Karls-Universit\"at T\"ubingen, D-72076 T\"ubingen, Germany}

\author{E. Lucioni}
\affiliation{Institut Non Lin\'eaire de Nice, CNRS, Universit\'e de Nice Sophia-Antipolis, 06560 Valbonne, France}
\affiliation{Dipartimento di Fisica, Universit\`a Degli Studi di Milano, Via Celoria 16, I-20133 Milano, Italy}

\author{Ph.W. Courteille}
\affiliation{Institut Non Lin\'eaire de Nice, CNRS, Universit\'e de Nice Sophia-Antipolis, 06560 Valbonne, France}
\affiliation{Physikalisches Institut, Eberhardt-Karls-Universit\"at T\"ubingen, D-72076 T\"ubingen, Germany}
\affiliation{Instituto de F\'isica de S\~ao Carlos, Universidade de S\~ao Paulo, 13560-970 S\~ao Carlos, SP, Brazil}

\author{N. Piovella}
\affiliation{Dipartimento di Fisica, Universit\`a Degli Studi di Milano, Via Celoria 16, I-20133 Milano, Italy}

\author{R. Kaiser}
\affiliation{Institut Non Lin\'eaire de Nice, CNRS, Universit\'e de Nice Sophia-Antipolis, 06560 Valbonne, France}

\date{\today}

\begin{abstract}
Cooperative scattering of light by an extended object such as an atomic ensemble or a dielectric sphere is fundamentally different from scattering from many pointlike scatterers such as single atoms. Homogeneous distributions tend to scatter cooperatively, whereas fluctuations of the density distribution increase the disorder and suppress cooperativity. In an atomic cloud, the amount of disorder can be tuned via the optical thickness, and its role can be studied via the radiation force exerted by the light on the atomic cloud. Monitoring cold $^{87}\text{Rb}$ atoms released from a magneto-optical trap, we present the first experimental signatures of radiation force reduction due to cooperative scattering. The results are in agreement with an analytical expression interpolating between the disorder and the cooperativity-dominated regimes.
\end{abstract}


\maketitle
Coherence effects in spontaneous radiation and scattering of light by small and extended clouds of atoms has been studied extensively in the past, starting with pioneering work by Dicke \cite{Dicke1954}.
In the context of cooperative effects in the scattering properties of a large cloud of atoms \cite{Scully_Science, Eberly06, Glauber08}
and localization of light with cold atoms \cite{AGK2008, Sokolov2009}, the interaction of one photon with a cloud of N atoms has seen renewed interest.
Since the development of Bose-Einstein condensates (BEC) with ultracold atoms, light scattering has been shown to be strongly modified in quantum degenerate systems of bosons and fermions \cite{Javanainen1994, Javanainen1995, Ketterle1999, Ketterle1999Science, Davidson2008}.
Using a simple quantum description of the photon scattering by $N$ cold atoms, we studied in a recent paper \cite{Courteille2009} the mechanical impact on cold atoms of cooperative scattering of light. We showed that the radiation force acting on a large cloud of atoms can be drastically reduced, due to both increased forward scattering and a reduced total scattering cross section.
Monitoring the modification of the atomic motion induced by light scattering, this study is complementary to the more commonly investigated features of the scattered light \cite{AGK2008, Sokolov2009} and may provide a new tool for the experimental investigation of cooperativity in the interaction between light and cold/ultracold matter, including subradiance and localization of light by disorder.

In this letter we report the first observation of cooperative scattering in a large atomic cloud where disorder due to density fluctuations tends to suppress cooperativity. The measured reduction of the radiation force is in very good agreement with the analytical expression derived in \cite{Courteille2009}. We discuss the role of the disorder in the radiation force before presenting
our experimental setup. The experimental results will then be compared to the theoretical predictions and we discuss further directions of this interesting line of research.

We consider a cloud of $N$ two level atoms (positions $\mathbf r_j$, transition wavelength $\lambda$, excited state lifetime $1/\Gamma$), excited by an incident laser  propagating along the direction $\hat{\mathbf e}_x$ (Rabi frequency $\Omega_0$, detuning $\Delta_0$, wavevector $\mathbf{k}_0$). We will restrict our analysis to the low intensity limit, neglecting non linear effects which could arise when more than one atom is simultaneously excited.
Using the Markov approximation, in which the radiation escapes from a large atomic cloud in a time much shorter than the characteristic decay time,
we derived the following steady-state expression for the average radiation force $F_c$ acting on the center of mass of large clouds of atoms \cite{Courteille2009}:
\begin{equation}\label{EqCooperativeforce}
 \frac{F_c}{F_1}=\frac{4\Delta_0^2+\Gamma^2}{4\Delta_0^2+N^2s_N^2\Gamma^2}~Ns_N\left[1-\frac{f_N}{s_N}\right],
\end{equation}
where $F_1 = \hbar k_0\frac{\Gamma}{2}\frac{\Omega_0^2/2}{\Delta_0^2+\Gamma^2/4}$ is the single atom radiation force.
From the structure function $S_N(\mathbf{k})=\frac{1}{N}\sum_{j=1}^N\mathrm{e}^{i\left(\mathbf{k}-\mathbf{k}_0\right)\cdot\mathbf{r}_j}$, we derived
the average value $s_N=\langle|S_N|^2\rangle_{\theta,\phi}$ and the phase function $f_N=\langle|S_N|^2\cos\theta\rangle_{\theta,\phi}$, where the average is taken over the total solid angle of emission of a photon with wavevector $\mathbf{k}$, at an angle $\theta$ with $\mathbf{k}_0$.
For smooth density distributions $n(\mathbf{r})$, one can compute the structure function by replacing the sum with an integral ($s_N\rightarrow s_\infty$, $f_N\rightarrow f_\infty$).
We found that $F_c$ can be influenced by different effects. On one side, the finite extent of the atomic cloud can produce strong forward oriented scattering, as in the case of Mie scattering or, more precisely, Rayleigh-Debye-Gans scattering \cite{Hulst1957}. The balance between the momentum of the incident and scattered photons and the atoms indicate that for forward emission, the net recoil imprinted onto the atoms is vanishing, resulting in a reduced radiation force ($f_\infty\approx s_\infty$).
A different contribution to the reduction of the radiation force can be seen in the prefactor of Eq. (\ref{EqCooperativeforce}), which would appear
even in a case of isotropic scattering (i.e. $f_\infty\approx0$). The importance of this prefactor can be understood
from the cooperative coupling of several atoms into the same vacuum mode \cite{AGK2008}. The number of available modes for large spherical clouds (with $n(r) \propto \mathrm e^{-r^2/(2 \sigma_R^2)}$) can be estimated by $N_{m}\propto~(k\sigma_R)^2$, resulting in an atom number per mode scaling as $N/N_{m} \propto ~ N/(k\sigma_R)^2$.
This scaling is conveniently related to the on-resonant optical thickness of the atomic cloud $b_0=(3\lambda^2/2\pi)\int \mathrm d z~n(0,0,z)=3N/(k\sigma_R)^2$. Using $s_\infty\approx1/4(k\sigma_R)^2$ and $f_\infty \approx s_\infty -2s_\infty^2$ \cite{Courteille2009}, one predicts for a smooth density distribution $Ns_\infty=b_0/12$ and thus the following cooperative radiation force:
\begin{equation}\label{EqSmoothCooperativeforce}
	 \frac{F_{c,\infty}}{F_1}=\frac{4\Delta_0^2+\Gamma^2}{4\Delta_0^2+\left(\frac{b_0}{12}\right)^2\Gamma^2}~\frac{b_0}{24(k\sigma_R)^2}.
\end{equation}
This expression has been derived assuming a smooth density distribution and is consistent with a vanishing radiation force for very large $b_0$. However, Eq.(\ref{EqSmoothCooperativeforce}) also predicts a vanishing radiation force for $b_0 \rightarrow 0$, where one expects single atom physics to become relevant (and $F_{c} \rightarrow F_1$).
\begin{figure}[b]
		\centerline{{\includegraphics[height=3.5cm]{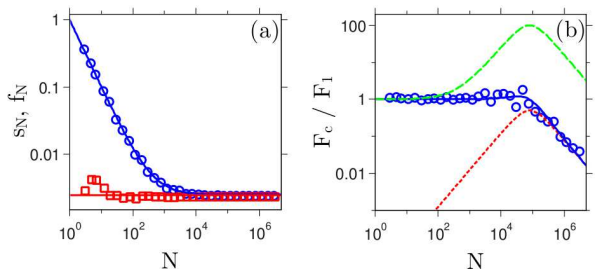}}}

		\caption{(color online) Comparison between analytical expressions and numerical evaluation for $k\sigma_R=10$ with a configuration average on 10 realizations as a function of atom number.
(a) Results for $s_N$ (blue circles), $f_N$ (red squares), and analytical expressions $1/N+s_\infty$ (blue line), $f_\infty$ (red line). (b) Forces acting on a cloud of atoms with $\Delta_0=-100~\Gamma$: numerical evaluation (blue circles) of the average cooperative force. The full lines indicate (i) the force in the presence of isotropic scattering, i.e. assuming $f_N=0$ (green dashed line), (ii) the force for continuous density distributions without disorder (red dotted line) and (iii) the total force taking into account cooperative scattering and disorder (blue line).}
		\label{Fig1}
	\end{figure}
This paradox can be solved by realizing that a continuous
density distribution cannot account for small atom numbers because it
passes by on the dominant role of disorder in the low density limit,
not untypical for many treatments of coherent scattering of light
\cite{Scully_Science, Eberly06, Glauber08, Javanainen1994, Javanainen1995}.
Replacing the sum by an integral in the evaluation of the structure function eliminates scattering from microscopic inhomogeneities in the cloud of atoms, reminiscent of a
coarse graining when computing effective indices. As in the comparison between cooperative scattering and disorder induced localization \cite{AGK2008}, one thus needs to estimate the fluctuations of $s_N$ and $f_N$ in Eq. (\ref{EqCooperativeforce}). One can show that to very good approximation [see Fig. \ref{Fig1}(a)], one has
\begin{equation}\label{EqsN}
s_N \approx\frac{1}{N}+s_\infty.
\end{equation}
The first term in $s_N$ describes scattering by a single atom, whereas the second term describes interferences between scattering from different atoms.
For $f_N$ we noticed that even though its configuration average is very close to $f_\infty$, the shot
to shot fluctuations of $f_N$ are bounded by $|f_N-f_\infty|<1/N$ and have a root mean square scaling as $\sigma_f\propto 1/N$.
It is now straightforward to estimate the average cooperative radiation force as a function of on-resonant optical thickness $b_0$ and size $\sigma_R$:
\begin{equation}\label{EqCooperativeforceNatom}
	 \frac{F_{c,N}}{F_1}=\frac{4\Delta_0^2+\Gamma^2}{4\Delta_0^2+\left(1+\frac{b_0}{12}\right)^2\Gamma^2}\left[1+\frac{b_0}{24(k\sigma_R)^2}\right].
\end{equation}
This expression describes the modification of cooperative scattering via its impact on the radiation force taking into account both, cooperative effects and the role of disorder. This expression allows us to make realistic estimates for an experimental implementation to measure the cooperative modification of the radiation force.
In Fig. \ref{Fig1}(b), we highlight the various contributions to the reduction of the radiation force for extended clouds of cold atoms by choosing $k\sigma_R=10$ and $\Delta_0=-100~\Gamma$.

\begin{figure}[b]
		\centerline{{\includegraphics[height=4.7cm]{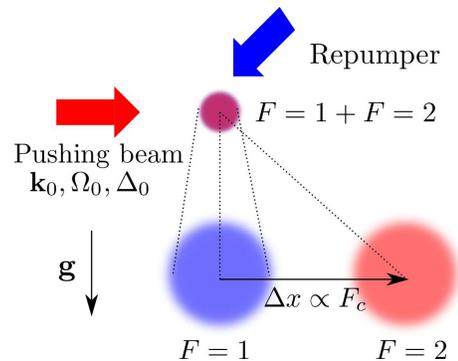}}}
		\caption{(color online) Experimental scheme to measure the cooperative radiation force. Atoms repumped into the $F=2$ state are pushed by a quasiresonant laser beam, whereas atoms remaining in $F=1$ only experience free fall.}
		\label{Fig2}
	\end{figure}

\begin{figure}[b]
		\centerline{{\includegraphics[height=6.5cm]{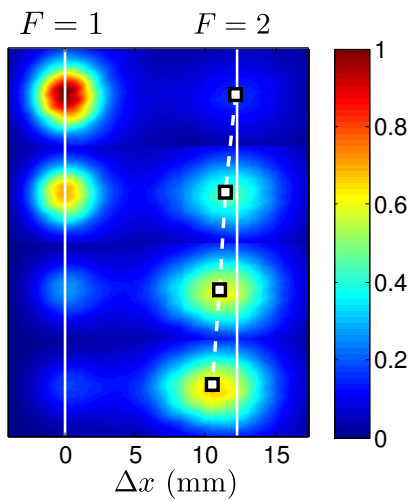}}}
		\caption{(color online) Fluorescence imaging (color-coded scale normalized to the maximum signal) of atom clouds in the $F=1$ (left) and in the $F=2$ (right) hyperfine ground state. The white squares locate the center of the $F=2$ cloud which is displaced by an amount depending on its optical thickness. For this experiment $\Delta_0 = -4.2 \, \Gamma$ and from top to bottom $b_0 = 5.5, 13.5, 17.9, 19.1$.}
		\label{Fig3}
	\end{figure}

Let us now describe the experimental setup and protocol to measure the cooperative radiation force (see Fig. \ref{Fig2}). Different regimes can be investigated, depending on the shape, size, optical thickness of the cloud and the laser detuning. In order to illustrate the cooperative scattering, we measure the radiation force by changing the optical thickness of the cloud, keeping all other parameters constant. This is done using the following procedure.
We first load a MOT of $^{87}\text{Rb}$ from a vapor cell.
All lasers are tuned close to the D2 line of $^{87}\text{Rb}$ and are derived from distributed feedback laser diodes, conveniently amplified with a tapered amplifier and controlled via acousto-optical modulators. Large laser beams and ultraviolet LEDs are used to optimize the loading of the MOT, where we trap $\approx 10^9$ atoms in $2$ s. We then apply a $50$ ms temporal dark MOT period where the intensity of the repumping laser is reduced by a factor of $10$ and the detuning of the cooling laser is increased from $-4 \, \Gamma$ to $-8 \, \Gamma$. This allows to compress the cloud and to produce a homogeneous Gaussian shaped distribution of atoms, mainly in the $F=1$ hyperfine ground state, with a temperature $T \approx 40 \, \mu\text{K}$. We then switch off all laser beams and magnetic field gradient, leaving the atoms in free fall. Hyperfine pumping into the $F=1$ state is completed during a $0.6$ ms molasses period without repumper. Before the cloud expands significantly, we apply a repump laser for $0.5$ ms. Changing the intensity of this repump laser allows us to prepare a controlled number of atoms in the $F=2$ hyperfine ground state without changing the volume of the cloud. We thus control the optical thickness $b_0$ of the almost Gaussian cloud with constant size $\sigma_R \approx 650 \, \mu \text{m}$ ($k \sigma_R \approx 5 \, 10^3$). We then apply an horizontal, circular polarized ``pushing" beam, tuned close to the $F=2\rightarrow F'=3$ transition for $0.8~\text{ms}$. We adjust the intensity depending on the detuning $\Delta_0$ in order to scatter around 100 photons per atom in the $F=2$ state, allowing for a net separation of the cloud displaced by the pushing beam  (see Fig. \ref{Fig3}).

Atoms in the $F=1$ state are not affected by this pushing beam.
After an additional time of flight (TOF) of $13\,\text{ms}$, we detect all atoms by complete repumping and fluorescence imaging at $90^\circ$ from the pushing beam. This experimental protocol allows us to distinguish atoms which have been exposed to the radiation force (in $F=2$) from those unaffected by the pushing beam (in $F=1$). The spatial displacement $\Delta x$ of the cloud after a fixed TOF is proportional to the average radiation force.

A typical sequence of observed images at $\Delta_0=-4.2~\Gamma$ for different repump intensities, and thus different optical thickness, is shown in Fig. \ref{Fig3}. An average of $100$ images have been taken to improve the signal to noise ratio, producing a slight broadening of the $F=2$ cloud due to intensity fluctuations of the pushing beam.
One can clearly distinguish the reduction of the spatial displacement, and hence of the radiation force, for an increasing number of atoms in $F=2$.

\begin{figure}[b]
		\centerline{{\includegraphics[height=5cm]{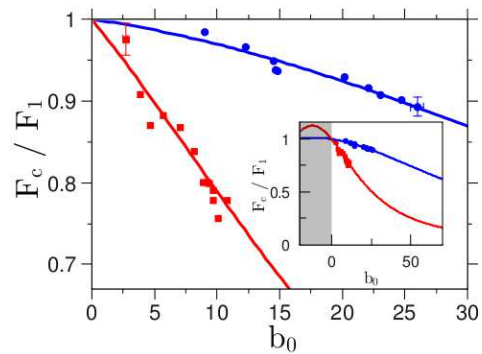}}}
		\caption{(color online) Experimental data and fits using the cooperative radiation force in the presence of disorder for $\Delta_0=-1.9 \, \Gamma$ (red squares) and $\Delta_0=-4.2 \, \Gamma$ (blue circles), with typical error bars from shot to shot fluctuations. The shadowed area in the inset corresponds to the nonphysical region $b_0 < 0$.}
		\label{Fig4}
	\end{figure}

A more quantitative comparison to the prediction of Eq. (\ref{EqCooperativeforceNatom}) for two different detunings is shown in Fig \ref{Fig4}, where we plot the radiation force,
normalized to unity for zero optical thickness, for two different values of the laser detuning : $\Delta_0=-1.9\,\Gamma,$ and $\Delta_0= -4.2\,\Gamma$. An ab intitio estimation of the radiation force from the measured values of the intensity, detuning and timing of the pushing beam yields a satisfactory agreement within $20\%$. The modified radiation force has a Lorentzian dependence on the optical thickness $b_0$ [see Eq. (\ref{EqCooperativeforceNatom})] with a maximum at $b_0=-12$.
Fitting the experimental data with the laser detuning as the only free parameter, we obtain for the two curves $\Delta_0^{fit} = -1.4 \pm 0.2\, \Gamma$ and $\Delta_0^{fit} = -4.3 \pm 0.2\, \Gamma$.
We have noticed that the good agreement with the theoretical model is less accurate (by up to $20\%$) if we use a linearly polarized pushing beam, because the Raman transitions between different Zeeman levels in the case of a linearly polarized beam are not included in our 2 level model. On the other hand, at a scattering rate of $\approx 100\, \text{kHz}$,
optical pumping into the $|F=2, m_F=2\rangle$ state by a circularly polarized beam
is only weakly affected by Larmor precession for residual magnetic fields below $20\,\text{mG}$.
For laser detunings too close to the atomic resonance, we observed an important deformation of the displaced atomic cloud, indicating the role of multiple scattering with different importance in the center and the edges of the atomic cloud. The above mentioned protocol to measure the cooperative radiation force in presence of disorder
is however restricted to the quasiresonant regime. Indeed, for larger laser detunings, optical hyperfine pumping puts more stringent constraints on the number of photons which can be scattered on the cycling $F=2\rightarrow F'=3$ transition. We have already seen hyperfine optical pumping which
produces a tail of atoms between the initial cloud, and the atoms in the $F=2$ state which are pushed by the quasiresonant laser beam. Different protocols can be used, but for a quantitative measurement of radiation force, it is very convenient to be able to change the number of interacting atoms without changing the volume of the atomic cloud. In previous experiments, we have observed a reduction of the radiation force for increasing atom numbers by using a red detuned dipole trap with a detuning in the range of $50-200$ GHz.
Even though strong qualitative evidence for the reduction of the radiation force has been observed, a quantitative comparison with our theoretical model has been difficult, partly due to the uncontrolled  shape of the elongated atomic cloud.

The good agreement between experiments and theory might be somewhat surprising, considering that in our model
we have assumed atoms at rest and that only one photon is present in the atomic cloud at the same time.
The experiments presented in this letter were performed at low intensity (saturation parameter $s<~10^{-2}$). However, as we couple to a complex structure (an extended cloud of atoms), it would be interesting to see at what light intensities deviations from our linear response model (``one photon") will appear. We obviously expect such deviations when entering the superradiant regime, which requires a large fraction of the cloud to be in the excited state at the same time.
Another important assumption is that we neglect any Doppler broadening which could lead to a dephasing between all induced dipoles \cite{Labeyrie2006}. Experiments on superradiant scattering performed in BECs \cite{Ketterle1999Science} have for instance noticed that due to such dephasing processes no signs of superradiance were observed above the transition temperature for BEC. In contrast to our experiments, such superradiance with escape rates scaling as $N^2$ requires a strong excitation with about half of the atoms excited at the same time \cite{Dicke1954}. Also, in our geometry, the cooperative scattering is strongly forward directed, reducing the impact of Doppler shift $(\mathbf{k}_0-\mathbf{k})\cdot\mathbf{v}$ compared to a geometry where cooperative scattering occurs at $90^\circ$ from the incident beam.

In conclusion, we have derived an analytical expression for the radiation force taking into account cooperative effects as well as disorder. This allows for a continuous description for the radiation force between microscopic, single atom Rayleigh scattering and macroscopic cooperative scattering.
We have experimentally observed this effect which manifests itself by a reduction of the radiation force with increasing atom number, well described by a simple formula taking account of the finite optical thickness of the cloud. The experiment thus represents the first measurement of the impact of cooperativity and disorder on radiation force.
Understanding how to connect disorder and cooperativity can be important for the study of localization of light in cold atoms \cite{AGK2008, Sokolov2009}. In this context the role of increased spatial densities will need to be considered, where the phase shift of the driving and scattered fields by the atomic cloud will become increasingly important. We also expect the effects studied in this letter to have sizable consequences on the heating rate in optical dipole traps.
One ingredient in the model is the driven timed Dicke state \cite{Courteille2009}, for which important entanglement properties have been predicted \cite{Eberly06, vanEnk2005}. As a classical description of many driven dipoles also yields strong forward scattering \cite{Hulst1957}, it would be interesting to design an entanglement witness \cite{Lewenstein2000} to probe the degree of relevant entanglement when N atoms are excited by a laser.

\acknowledgments{This work was supported by ANR CAROL (project ANR-06-BLAN-0096) and by the Deutsche Forschungsgemeinschaft (DFG) under Contract No. Co~229/3-1. S.B. acknowledges a grant from DAAD and E.L. support from INTERCAN.}

\end{document}